\documentclass[a4paper,11pt]{article}

\usepackage[top=3cm,bottom=3cm,left=2cm,right=1.5cm,nohead,includefoot,foot=0.6cm,footskip=1.099cm]{geometry}

\usepackage[ascii]{inputenc}
\usepackage[T1]{fontenc}
\usepackage[english,english]{babel}
\usepackage{amsmath}
\usepackage{amssymb,amsfonts,textcomp}

\begin{document}


\title{Rigid covariance, equivalence principle and Fermi rigid coordinates: gravitational waves}

\author{Xavier Ja\'{e}n\thanks{Dept. de F\'{i}sica, Universitat Polit\`{e}nica de Catalunya, Barcelona, Spain, e-mail address: xavier.jaen@upc.edu}} 

\date{}

\maketitle


\begin{abstract}
For a given space-time and for an arbitrary time-like geodesic, we analyze the conditions for the construction of Fermi coordinates so that they are also rigid covariant. We then apply these conditions to linear plane gravitational waves.
\end{abstract}

\textit{Keywords:} Rigid motion, Fermi coordinates, Equivalence principle, Linear plane gravitational wave

\section{Introduction}\label{s_1}
In a series of recent papers, \cite{01-jaen_1,02-jaen_2,03-jaen_3,04-jaen_4}, we have presented a formulation of the general theory of relativity with a covariance group that is smaller than usual. General covariance implies having ten potentials to describe gravitation; but four of them can be eliminated by means of coordinate transformations. So, only six potentials are really necessary. In those papers we have reduced the covariance group to bring it as close as possible to the usual group of rigid motions, in a formulation with six potentials.
It is not a novelty to use the concept of rigidity in relativity. In 1909 Born naturally extended this concept to relativity, but it turned out to be inconsistent in too many cases \cite{Born1909}. In spite of the difficulties, some authors \cite{Bona83, Bel90, Bel96, Llosa04, Coll07} think that rigidity could be the missing piece so that relativity acquires a state of maturity comparable to Newtonian mechanics. The differences between these authors are related to the difficulty in carrying out this natural extension.

In some aspects, our rigid formulation of general relativity is remarkably close to the corresponding formulation of Newtonian gravitation. This allows us to analyze the topics mentioned in the title of this paper. We will see that, at the Newtonian level, there are gauge transformations that, although they do not entail any relevant physical changes, allows us to implement a Newtonian equivalence principle, in which rigid motions are those that lead us to the locally inertial reference system, where the effects of gravitation are locally canceled out. 

When we analyze the same topic in general relativity we see how new coordinate transformations appear that we can still say are rigid in the sense that they leave the space-time metric shape invariant. We will see how we can implement the equivalence principle using these transformations together with the gauge transformations, already seen at the Newtonian level, which in general relativity induce temporal transformations.

As a consequence of this study, we can conjecture an implementation of the (relativistic) equivalence principle where rigid motions play a leading role that is very close to the Newtonian case. We will see how the theoretical framework that arises can be applied to some relevant space-times; the most interesting being that of linear plane gravitational waves. These have already been analyzed in \cite{04-jaen_4}, where we found a set of rigid reference systems with which to express the metric. In the present study, we will see how we can choose a system from that previously determined set that defines a rigid locally inertial system. In other words, for linear plane gravitational waves, we will find rigid coordinates which in turn are Fermi coordinates attached to a geodesic.

In 1922 Fermi \cite{05-fermi} defined the so-called Fermi coordinates (FC), a construction of a coordinate system
%
%
\footnote{Greek indices are used for space-time $\mu ,\nu  = 0,1,2,3$. $	{\eta _{\mu \nu }} = {\rm{diag}}( - 1,1,1,1)$. Latin indices are used for space $i,j = 1,2,3$. We use the usual Euclidean vector notation: scalar products, ``$\cdot$`` , norms, $x \equiv \left| {\vec x} \right|$, vector products ``$ \times $`` and the operation of raising and lowering indices are always performed using the Euclidean metric with Euclidean components ${\delta _{ij}}$ and the corresponding totally antisymmetric volume form ${\delta _{ijk}}$ with
${\delta _{123}}=1$. }
%
%
${x^\mu } = \left\{ {t,\vec x} \right\}$, that near a given time like curve, $\vec x=0$, the space-time looks like a flat one, that is
\begin{equation}\label{eq_0}
{\left. {d{s^2}} \right|_{\vec x = 0}} = {\eta _{\mu \nu }}d{x^\mu }d{x^\nu }\;\;,{\left. {\Gamma _{\mu \nu }^\sigma }
\right|_{\vec x = 0}} = 0
\end{equation}
which is equivalent to:
%
%
\footnote{The notation $A + O\left( {{x^n}} \right)$ means that $O\left( {{x^n}} \right)$ is an object of the same kind as $A$ (scalar, vector, tensor, etc.) whose components in the natural basis associated with $\vec x$ are of the order ${\left| {\vec x} \right|^n}$ or higher.}
%
%
\begin{equation}
d{s^2} = {\eta _{\mu \nu }}d{x^\mu }d{x^\nu } + O\left( {{x^2}} \right)
\end{equation}

Manasse and Misner \cite{06-manasse} specialized Fermi's coordinates for a given time-like geodesic and, in order to find a physical meaning for Fermi's proposal, they chose a particular set of coordinates from those satisfying (\ref{eq_0}), which they called Fermi normal coordinates (FNC). The name was suggested after Riemann \cite{07-riemann} introduced the so-called Riemann normal coordinates, any coordinate system $\left\{ {t ,\vec x} \right\}$ for which the metric can be written in a neighborhood of a given space-time point $P$ as: 
\begin{equation*}
d{s^2} = \left\{ {{\eta _{\mu \nu }} + \frac{1}{3}{{\left. {{R_{\alpha \beta \mu \nu }}} \right|}_P}{x^\alpha }{x^\beta
}} \right\}d{x^\mu }d{x^\nu } + O\left( {{{({x^\mu })}^3}} \right)
\end{equation*}
where ${{R_{\alpha \beta \mu \nu }}}$ is the Riemann tensor and $O\left( {{{({x^\mu })}^3}} \right)$ labels the terms of third order or higher in the space-time coordinates. Following the Fermi and Riemann construction, Manasse and Misner showed that in their FNC, the metric coefficients $g_{\mu\nu}$ fulfill:

\begin{equation}\label{eq_2}
\begin{array}{*{20}{l}}
{{g_{00}} =  - 1 + {{\left. {{R_{0\ell 0m}}} \right|}_{\vec x = 0}}\;{x^\ell }{x^m} + O\left( {{x^3}} \right)}\\
{{g_{0i}} = \frac{2}{3}{{\left. {{R_{0\ell im}}} \right|}_{\vec x = 0}}\;{x^\ell }{x^m} + O\left( {{x^3}} \right)}\\
{{g_{ij}} = {\delta _{ij}} + \frac{1}{3}{{\left. {{R_{i\ell jm}}} \right|}_{\vec x = 0}}\;{x^\ell }{x^m} + O\left(
{{x^3}} \right)}
\end{array}
\end{equation}
Ni and Zimmermann  \cite{08-ni} generalized the work of Manasse and Misner for any initial time-like non-geodesic world-lines and included arbitrary rotation. They studied accelerated reference systems with rotation. Li and Ni in  \cite{09-li_1,10-li_2}, Nesterov  \cite{11-nesterov}, Marzlin \cite{12-marzlin_1}, have all made contributions in the same direction, calculating increasing approximation orders of the metric using FNC.

In the present work, our focus is on the way FNC are constructed. Manasse and Misner based FNC on geodesics in order to be close to a reasonable interpretation. From a geometrical point of view, it is totally understandable to use geodesics. However, the physical meaning is not satisfactory, or at least it does not seem entirely clear to all authors; see, for example, the comments in  \cite{13-rakhmanov,14-marzlin_2} and especially in  \cite{15-delva}. 

In this context, it seems reasonable to look for Fermi-like coordinates based on some alternative procedure that can be compared with the previous constructions. The idea of not basing the spatial Fermi coordinates on the geodesic distance is not new  \cite{14-marzlin_2,15-delva}. 

As we have a rigid covariant formulation of general relativity, it seems reasonable to try to use the rigid concept and the rigid covariance in order to find FC that are as close as possible to rigid ones. We will see that this approach is fruitful and we will call the resultant coordinates Fermi rigid coordinates (FRC). Of course, we demand that these coordinates satisfy (\ref{eq_0}) near the geodesic $x^i=0$ but not necessarily   (\ref{eq_2}). Up to first order, the only difference between FNC and FRC will be the way they are constructed.

In  \cite{04-jaen_4}, we already derived a rigid formulation for linear plane gravitational waves. In fact, we found a broad set of possible rigid coordinates without having a physical criterion for choosing one of them. The present work can therefore be seen as the continuation of  \cite{04-jaen_4}. This means that our purpose is not to reach a high order in the calculation. Rather, it is to prove that the concept of rigidity can be useful in the construction of reference systems and coordinates with physical significance, at least to the order in which physical significance can be given, and this means that we will explore the framework of Fermi's coordinates.

The paper is organized as follows: Section $\S$\ref{s_2} is devoted to reviewing the rigid covariant formulation of Newtonian gravitation. In Section \ref{s_3}$\S$, we look at the gauge invariance of this formulation; while in Section $\S$\ref{s_4}, we offer a complete version for the implementation of the equivalence principle for Newtonian gravitation. In Section $\S$\ref{s_5}, within the context of general relativity, we study Painlev\'{e}-Gullstrand space-times for which we give the conditions for FRC and we find them up to first order in the metric. Then, in Section $\S$\ref{s_6}, we deal with a general space-time in the covariant rigid formulation. We analyze, beyond the usual rigid motions, the transformations that leave the metric shape invariant and define what we mean by an isochronous geodesic congruence with respect to a given geodesic. In Section $\S$\ref{s_7}, we use what we have learned in the previous section and present a protocol for finding FRC that we solve completely up to first order in the metric. Finally, in Section $\S$\ref{s_8}, we apply what we have learned to find FRC for linear plane gravitational waves.

\section{Newtonian gravitation}\label{s_2}
Over the years and for various reasons several authors have been interested in the structure of the Maxwell-Lorentz equations beyond their application to electromagnetism. One such example of this interest that particularly fascinates us is that expressed by R. Feynmann and reported by F.J. Dyson  \cite{16-dyson}. We are interested in a particular family of gauge transformations that are especially suited to gravity. Let us look at them first in a simple and generic way, and then we will consider the more specific gravitational case. 

Given a Lagrangian for a particle of the form: 
\begin{equation}\label{eq_3}
L = \frac{1}{2}{\dot {\vec x}^2} + \vec A\cdot\dot {\vec x} + B
\end{equation}
where $\vec A$ i $B$ are functions of $(\vec x, t)$, we can always write it in the form: 
\begin{equation}\label{eq_4}
L = \frac{1}{2}{\left( {\dot {\vec x} - \vec V} \right)^2} + \frac{{d\tau }}{{dt}}
\end{equation}
where $\vec V$ and $\tau$ are functions of $(\vec x, t)$. The proof is very simple. From the equality between   (\ref{eq_3}) and   (\ref{eq_4}), we get the conditions: 
\begin{equation}\label{eq_5}
\vec V =  - \vec A + {\partial _{\vec x}}\tau 
\end{equation}
and:
\begin{equation}\label{eq_6}
{\partial _t}\tau  = B - \frac{1}{2}{\left( {\vec A - {\partial _{\vec x}}\tau } \right)^2}
\end{equation}
What is not so well known, is that if $H$ is the Hamiltonian associated with  (\ref{eq_3}),   (\ref{eq_6}) is equivalent to the Hamilton-Jacobi equation with ${\tau }$ playing the role of the action, ${\partial _t}\tau  + H(\vec x,\vec p = {\partial _{\vec x}}\tau ,t) = 0$, and  (\ref{eq_5}) defines the velocity field, which is equivalent to $\vec V =\frac{{\partial H}}{{\partial \vec p}}(\vec x,\vec p = {\partial _{\vec x}}\tau ,t)$. For each ${\tau }$ solution of the Hamilton-Jacobi equation  (\ref{eq_6}), we have a potential $\vec V$ from  (\ref{eq_5}) and, with it, a Lagrangian of the form   (\ref{eq_4}).

The equations of motion corresponding to  (\ref{eq_4}) can be written: 
\begin{equation}\label{eq_7}
\ddot {\vec x} = \vec g + \dot {\vec x} \times \vec \beta 
\end{equation}
where the relationships between the fields, ${\vec g}$ and ${\vec \beta }$, and the potential ${\vec V}$ are:
\begin{equation}\label{eq_8}
\begin{array}{*{20}{l}}
{\vec g = {\partial _{\vec x}}\left( {\frac{{{V^2}}}{2}} \right) + {\partial _t}\vec V}\\
{\vec \beta  =  - {\partial _{\vec x}} \times \vec V}
\end{array}
\end{equation}

This result is applicable to gravitation. As we saw in  \cite{01-jaen_1}, in any
rigid reference system $S$, the gravitational field can be represented by a potential $\vec V(\vec x,t )$ where $\left\{ {t ,\vec x} \right\}$ are the rigid time and space Euclidean coordinates. The equation of motion is of the Lorentz type   (\ref{eq_7}) and   (\ref{eq_8}). We will call a geodesic trajectory of the Newtonian gravitational field   (\ref{eq_8}) any solution of the equation of motion   (\ref{eq_7}).  The gravitational field equations for ${\vec V}$ can be written in terms of ${\vec g}$ and ${\vec \beta}$ in the form: 
\begin{equation}\label{eq_9}
\begin{array}{*{20}{l}}
{{\partial _{\vec x}}\vec g - \frac{1}{2}{\beta ^2} =  - 4\pi G\rho }\\
{{\partial _{\vec x}} \times \vec \beta  = 0}
\end{array}
\end{equation}
Under a rigid motion transformation, ${\vec x = \vec R + \vec y}$, we will go from the reference system
$S$, $\left\{ {t ,\vec x} \right\}$, to the reference system $S'$, $\left\{ {t ,\vec y} \right\}$. A translation will be represented by the vector $\vec R \equiv {R^m}(t)\;{\hat X_m}$, and a rotation by the matrix $R_i^m(t)$, so that ${\hat Y_i} = R_i^m(t){\hat X_m}$ where ${{{\hat X}_i}}$ and ${{{\hat Y}_i}}$ are the orthonormal basis of $S$ and $S'$ respectively. In the $S'$ system, the equations of motion and the field equations have the same shape as in the $S$ system, but now the gravitational potential is $\vec V'$, and the relation with ${\vec V}$ is given by:
%
\footnote{The total derivative of a scalar with respect to time will be indicated as $\dot {f} $. In $S$, with $\vec x = {x^m}{\hat X_m}$, we will use the notation $\dot {\vec x} \equiv {\dot {x}^m}{\hat X_m}$ and $\ddot {\vec x} \equiv {\ddot {x}^m}{\hat X_m}$; in $S'$, with $\vec y = {y^m}{\hat Y_m}$, $\dot {\vec y} \equiv {\dot {y}^m}{\hat Y_m}$ and $\ddot {\vec y} \equiv {\ddot {y}^m}{\hat Y_m}$.}
%
%
\begin{equation}\label{eq_10}
\vec V = \dot {\vec R} + \vec V' + \vec \Omega  \times \vec y
\end{equation}
where $\dot {\vec R} \equiv {\dot {R}^m}{\hat X_m}$ and $\vec \Omega  \equiv \frac{1}{2}\sum\limits_s {R_s^m\dot {R}_s^n{\delta _{mn}}^i} {\hat X_i}$. That is, the potential of the gravitational field transforms as a velocity field. 

The equation of motion  (\ref{eq_7}) includes pure gravitational effects in addition to Coriolis, centrifugal and Euler forces, and those due to translation. The inertial forces are due to inertial fields that are source-free solutions of the field equations   (\ref{eq_9}).

The trajectories that are solutions of $\dot {\vec x} = \vec V$ are solutions of the equation of motion   (\ref{eq_7}). That is, the gravitational potential ${\vec V}$ is itself a geodesic congruence.

The equation of motion   (\ref{eq_7}) admits a very simple Lagrangian formulation, already suggested at the beginning of this section, with a scalar covariant lagrangian under rigid motion transformations:
\begin{equation}
L = \frac{1}{2}{\left( {\dot {\vec x} - \vec V} \right)^2} = \frac{1}{2}{\left( {\dot {\vec y} - \vec V'} \right)^2}
\end{equation}

\section{Gauge transformations in Newtonian gravitation} \label{s_3}
Because the physical fields are ${\vec g}$ and ${\vec \beta }$, we can use any potential ${\vec V^*}$, instead of ${\vec V}$, that gives, via the same relationships, the same physical fields. An elegant way to analyze a gauge transformation is that described in \cite{03-jaen_3}, where the physical meaning of the potential ${\vec V}$ as geodesic congruence plays an important role, which means that gauge invariance becomes a requirement: if the potential ${\vec V}$ is a geodesic congruence, 
it seems quite natural to require that we can use any other geodesic congruence 
${\vec V^*}$ as a potential. This requirement is met: given the Lagrangian $L = \frac{1}{2}\left( {\dot {\vec x} - \vec V} \right)^2$ and the corresponding Hamiltonian $H$, we define the gauge transformation $\vec V \to {\vec V^*}$ through the relations: 
\begin{equation}\label{eq_12}
{\vec V^*} \equiv \frac{{\partial H}}{{\partial \vec p}}(\vec x,\vec p = {\partial _{\vec x}}{\tau ^*},t)
\end{equation}
where ${\tau ^*}$, the action, is any solution of the Hamilton-Jacobi equation:
\begin{equation}\label{eq_13}
{\partial _t}{\tau ^*} + H(\vec x,\vec p = {\partial _{\vec x}}{\tau ^*},t) = 0
\end{equation}
Under these conditions, it is easy to see that:
\begin{equation}
{L^*} + \frac{{d{\tau ^*}}}{{dt}} = \frac{1}{2}{\left( {\dot {\vec x} - {{\vec V}^*}} \right)^2} + \frac{{d{\tau
^*}}}{{dt}} = \frac{1}{2}{\left( {\dot {\vec x} - \vec V} \right)^2} = L
\end{equation}
$L^*$ and $L$ differ by a total derivative. Using the Lagrangian $L^*$ we obtain the same equation of motion as we obtain with $L$. The equation of motion associated with ${\vec V^*}$ is the same as that associated with ${\vec V}$. The conclusion is that the fields ${\vec g}$ and ${\vec \beta }$ are gauge invariant under the transformation $\vec V \to {\vec V^*}$.

\section{Newtonian equivalence principle}\label{s_4}
In  \cite{01-jaen_1}, we saw an unfinished version of the implementation of the Newtonian equivalence principle. Here we complete that implementation and emphasize some of its characteristics that will be useful at the relativistic level.

Given a gravitational potential ${\vec V}$ in a rigid reference system $S$, $\left\{ {t ,\vec x} \right\}$, for each geodesic $G$ , $\vec x(t)$, of the geodesic congruence ${\vec V}$ we define a rigid reference system $S'$,$\left\{ {t ,\vec y} \right\}$, as that which is rigidly related to $S$ according to the rigid transformation:
\begin{equation}
{\vec x = \vec x(t) + \vec y\;}
\end{equation}
where ${\vec x(t) = {x^m}(t){\mkern 1mu} {{\hat X}_m}}$ is the geodesic $G$, ${\vec y = {y^m}\,{{\hat Y}_m}}$, and ${{{\hat Y}_i} = R_i^m(t){\mkern 1mu} \,{{\hat X}_m}}$ with $R_i^m(t)$ being any matrix rotation solution of the equation: 
\begin{equation}\label{eq_16}
\vec \Omega (t) \equiv \sum\limits_s {R_s^m\dot {R}_s^n{\delta _{mn}}^i} {{\hat X}_i} = {\left[ {{\partial _{\vec x}}
\times \vec V} \right]_{\vec x = \vec x(t)}}
\end{equation}
That is to say, the origin of the reference system $S'$ moves along the geodesic trajectory $G$, $\vec x(t)$, belonging to the congruence ${\vec V}$. The rotation of $S'$, with respect to $S$, $\vec \Omega (t)$, is equal to the vorticity of the congruence ${\vec V}$ evaluated on the geodesic $G$.

With these conditions, the transformed potential, $\vec V'$, is, according to  (\ref{eq_10}) expressed in $\vec y$ coordinates and developed up to order $y$: 
\begin{equation}\label{eq_17}
\vec V'\left( {\vec y,t} \right) = {\left[ {\left( {\vec y\cdot{\partial _{\vec x}}} \right)\vec V + \frac{1}{2}\vec y
\times \left( {{\partial _{\vec x}} \times \vec V} \right)} \right]_{\vec x = \vec x(t)}} + O\left( {{y^2}} \right)
\end{equation}
In this expression, we make extensive use of Euclidean vector notation. To expand it, it is necessary to keep in mind the relationship between the basis of $S$ and $S'$, ${\hat Y_i} = R_i^m{\hat X_m}$, when resolving some products. For example, in ${\vec y\cdot{\partial _{\vec x}}}$, the two factors must be expressed in their defining basis and the change must be made. The notation is consistent because the rotation $R_i^j$ depends only on $t$. 

A direct calculation, using  (\ref{eq_17}), shows that the gravitational fields in the reference system $S'$, $\vec g'$ and $\vec\beta '$, that we obtain from the potential $\vec V'$, are null at the origin of $S'$:
\begin{equation}\label{eq_18}
\begin{array}{*{20}{l}}
{\vec g'\left( {\vec y = 0,t} \right) = {{\left[ {{\partial _{\vec y}}\left( {\frac{{{{V'}^2}}}{2}} \right) + {\partial
_t}\vec V'} \right]}_{\vec y = 0}} = 0}\\
{\;\vec \beta '\left( {\vec y = 0,t} \right) =  - {{\left[ {{\partial _{\vec y}} \times \vec V'} \right]}_{\vec y = 0}}
= 0}
\end{array}
\end{equation}
So, $S'$ is actually a rigid locally inertial reference system attached to $G$: $S_{IG}$.

This result is in accordance with the equivalence principle. If we take into account gauge transformations $\vec V \to{\vec V^*}$ we can extend this result to any reference systems $S'$ attached on an arbitrary geodesic, not only a geodesic solutions of $\dot {\vec x} = \vec V$. In this context, the implementation of the Newtonian equivalence principle can be stated thus:

\begin{quote}
\textit{\textbf{Newtonian equivalence principle}: if, in a rigid reference system $S$, we have a gravitational field, for any geodesic $G$ we can always associate a geodesic congruence $\vec V$ and use it to build a rigid locally inertial reference system $S_{IG}$ such that its origin moves along the geodesic $G$ and rotates according to the vorticity of $\vec V$ evaluated at $G$. At the origin of $S_{IG}$, the gravitational field is null, $\vec g'\left( {0,t } \right) = 0\;$ and $\vec \beta '\left( {0,t } \right) = 0$, and consequently, test particles move freely.}
\end{quote}

It should be noted that it is the gauge invariance of the theory that allows it to be in accordance with the equivalence principle. Moreover, it is remarkable that, given a geodesic $G$, there may be more than one associated congruence $\vec V$. The vorticity of these congruences, due to the gauge invariance, will be equal and therefore all of them will define the same rotation $\vec \Omega $ (see  (\ref{eq_16}) ), that is, the same rigid locally inertial reference system $S_{IG}$. 

We can give a Lagrangian version of this result. What is going to be interesting now is not the result but how we arrive at it. We start in the rigid reference system $S$, with the Lagrangian $L = \frac{1}{2}{\left( {\dot{\vec x} - \vec V} \right)^2}$ where, using the gauge freedom, we take $\vec V$ in such a way that it include the given geodesic $G$ to build the system $S_{IG}$. The Lagrangian in $S_{IG}$, with coordinates $\vec y$, can be written, taking into account expression (\ref{eq_17}) for $\vec V'$ and developing it around $\vec y = 0$, in the form:
\begin{equation}\label{eq_19}
L = \frac{1}{2}{\left( {\dot {\vec y} - \vec V'} \right)^2} = \frac{1}{2}{\dot {\vec y}^2} - \frac{{dF}}{{dt}} + O\left(
{{y^2}} \right)
\end{equation}
where $F = \frac{1}{2}\vec y\cdot\vec V' $. The equation of motion related to  (\ref{eq_19}) at the origin of the system $S_{IG}$, $\vec y = 0$, is $\ddot {\vec y} = 0$.

It is worth commenting that the gauge transformations  (\ref{eq_12}) and   (\ref{eq_13}) have the effect of adding a total derivative of a function to the original Lagrangian, which that turns out to be the action. This suggests that we can proceed in an alternative way to reproduce  (\ref{eq_19}). Let us start with the Lagrangian $L = \frac{1}{2}{\left( {\dot {\vec y} - \vec V'} \right)^2}$ and look for a gauge transformation $\vec V' \to {\vec V'^*}$, described by  (\ref{eq_12}) and   (\ref{eq_13}), but now using $\vec y$ coordinates, which conforms to ${\vec V'^*} = O\left( {{y^2}} \right)$. The result is, as expected, ${\tau ^*} =  - \frac{1}{2}\vec y\cdot\vec V'$, and we reproduce expression  (\ref{eq_19}).

We should point out here that at the Newtonian level, the action ${\tau ^*}$ does not have any relevant meaning.

\section{ Painlev\'e-Gullstrand space-times}\label{s_5}
Let us now turn to general relativity. We will begin by studying a space-time that in the rigid reference system $S$ has a Painlev\'e-Gullstrand metric form  \cite{01-jaen_1}, based on the form of the metric that Painlev\'{e}  \cite{17-painleve} and, independently, Gullstrand  \cite{18-gullstrand} proposed for Schwarzschild space-time. On $S$, we will take the space-time coordinates $\left\{ {\lambda ,\vec x} \right\}$ and, from now on in this paper, we will take $c=1$. The metric is thus:
\begin{equation}\label{eq_20}
d{s^2} =  - d{\lambda ^2} + {\left( {d\vec x - \vec Vd\lambda } \right)^2}
\end{equation}
where $\vec V$ is a function of $(\vec x, \lambda)$. If we adapt the factor so that $Ld\lambda $ becomes the proper time, the associated Lagrangian, $L$, is:
\begin{equation}\label{eq_21}
L = \sqrt {1 - {{\left( {\dot {\vec x} - \vec V} \right)}^2}} 
\end{equation}
These space-times are rigid covariants, defining rigid transformations exactly in the same way as in the Newtonian case, only now using the time $\lambda$. The potential ${\vec V}$ again transforms as a velocity (the relation (\ref{eq_10}) remains valid).

As we have seen in  \cite{03-jaen_3}, the metric  (\ref{eq_20}) belongs to a broad class of rigid covariant metrics of the form:
\begin{equation}\label{eq_22}
d{s^2} =  - d{\tau ^2} + \left( {{\tau _{,{x^i}}}{\tau _{,{x^j}}} + {\delta _{ij}}} \right)(d{x^i} - {V^i}d\lambda
)(d{x^j} - {V^j}d\lambda )
\end{equation}
where $\tau  = \tau \left( {\vec x,\lambda } \right)$. For the metric  (\ref{eq_20}), $\tau \left( {\vec x,\lambda } \right) = \lambda $. 

From  (\ref{eq_22}), it is easy to see that $U = {\partial _\lambda } + {V^i}\left( {\vec x,\lambda } \right){\partial_{{x^i}}}$ is a time-like geodesic congruence with proper time $\tau \left( {\vec x,\lambda } \right)$. 

The metric (\ref{eq_20}), as a member of the metric class (\ref{eq_22}), is invariant under gauge transformations $\left\{ \tau  = \lambda ,\vec V \right\}\to\left\{ {\tau ^*},{{\vec V}^*} \right\}$ as described in  \cite{03-jaen_3}. The difference with the Newtonian case is that now the action ${\tau ^*}$, as a solution of the Hamilton-Jacobi equation, is also a potential. This potential has its own meaning as the proper time of the geodesic congruence $U^*$.

If $H$ is the Hamiltonian associated with the Lagrangian  (\ref{eq_21}), the gauge transformation $\left\{ {\tau  = \lambda ,\vec V} \right\} \to \left\{ {{\tau ^*},{{\vec V}^*}} \right\}$ is defined in a similar way to the Newtonian case:
\begin{equation}
{\vec V^*} \equiv \frac{{\partial H}}{{\partial \vec p}}(\vec x,\vec p = {\partial _{\vec x}}{\tau ^*},\lambda )
\end{equation}
and ${\tau ^*}$ is any solution of the Hamilton-Jacobi equation: 
\begin{equation}
{\partial _\lambda }{\tau ^*} + H(\vec x,\vec p = {\partial _{\vec x}}{\tau ^*},\lambda ) = 0
\end{equation}

For each geodesic $G$, ${\vec x}(\lambda)$, of the congruence $\vec V$ of   (\ref{eq_20}), we can define a reference system $S'$ by performing the same rigid transformation, $\vec x\to\vec y$, as we did in the Newtonian equivalence principle in Section $
\S$\ref{s_4}. The result is:
\begin{equation}\label{eq_25}
d{s^2} =  - d{\lambda ^2} + {\left( {d\vec y - \vec V'\;d\lambda } \right)^2}
\end{equation}
where $\vec V'\left( {\vec y,\lambda } \right)$, as in the Newtonian case, can be expressed in
$\vec y$ coordinates, developed up to order $y$ and giving the same result as in 
(\ref{eq_17}).

Now we look for a gauge transformation of the form $\left\{ {\tau  = \lambda ,\vec V'} \right\} \to \left\{ \tau^*,\vec {V'}^* = O\left( y^2 \right) \right\}$. The Hamiltonian associated with (\ref{eq_25}) is:
\begin{equation}
H\left( {\vec y,\vec p,\lambda } \right) = \vec V'\cdot\vec p - \sqrt {1 + {p^2}} 
\end{equation}
and the Hamilton-Jacobi equation: 
\begin{equation}\label{eq_27}
{\partial _\lambda }{\tau ^*} + \vec V'\cdot{\partial _{\vec y}}{\tau ^*} - \sqrt {1 + {{\left( {{\partial _{\vec
y}}{\tau ^*}} \right)}^2}}  = 0
\end{equation}
We look for ${\tau ^*}$ solutions fulfilling: 

\begin{equation}\label{eq_28}
{\vec V'^*} = O\left( {{y^2}} \right) = \frac{{\partial H}}{{\partial \vec p}}(\vec y,\vec p = {\partial _{\vec y}}{\tau
^*},\lambda )
\end{equation}
If we write (\ref{eq_28}) except for the $O\left( {{y^2}} \right)$ terms, we have: 
\begin{equation}\label{eq_29}
\frac{{{\partial _{\vec y}}{\tau ^*}}}{{\sqrt {1 + {{\left( {{\partial _{\vec y}}{\tau ^*}} \right)}^2}} }} = \vec V'
\end{equation}
As $\vec V'$ is $O\left( y \right)$,   (\ref{eq_29}) means that ${\partial _{\vec y}}{\tau ^*} = O(y)$. So we can write   (\ref{eq_27}) and   (\ref{eq_29}) as:
\begin{equation}\label{eq_30}
\begin{array}{*{20}{l}}
{{\partial _\lambda }{\tau ^*} = 1 + O({y^2})}\\
{{\partial _{\vec y}}{\tau ^*} = \vec V' + O({y^2})}
\end{array}
\end{equation}
The integrability conditions are guaranteed by   (\ref{eq_18}), that is, ${\partial _{\vec y}} \times \vec V' = O({y^2})$, and by the fact that we are neglecting terms of order $O({y^2})$. We can check by direct calculation, in which expression (\ref{eq_17}) for $\vec V'$ is used, that a solution satisfying   (\ref{eq_30}) is:
\begin{equation}\label{eq_31}
{\tau ^*} = \lambda  + \frac{1}{2}\vec y\cdot\vec V' + O({y^3})
\end{equation}
which is the relativistic version of the function $F$ in the Newtonian case: expression   (\ref{eq_19}). We now define the time coordinate $t = {\tau ^*}(\vec y,\lambda )$. The explicit expression of $t$ can be found using   (\ref{eq_31}) and (\ref{eq_17}):
\begin{equation}\label{eq_32}
t = \lambda  + \frac{1}{2}\vec y\cdot{\left[ {\left( {\vec y\cdot{\partial _{\vec x}}} \right)\vec V + \frac{1}{2}\vec y
\times \left( {{\partial _{\vec x}} \times \vec V} \right)} \right]_{\vec x = \vec R}} + O\left( {{y^3}} \right)
\end{equation}
The metric in $\left\{ {t,\vec y} \right\}$ coordinates takes the form: 
\begin{equation}\label{eq_33}
d{s^2} =  - d{t^2} + {\delta _{ij}}d{y^i}d{y^j} + O({y^2})
\end{equation}
that is, the $\left\{ {t,\vec y} \right\}$ coordinates are FC and the system $S'$ is a rigid locally inertial reference system attached to the geodesic $G$, $S_{IG}$. These coordinates are those we announce in the Introduction above and name: Fermi rigid coordinates (FRC).

We should note that of the coordinates $\left\{ {t,\vec y} \right\}$ only $\vec y$ can be labeled as \emph{rigid} at any order, in the sense that the $\left\{ {\lambda ,\vec y} \right\}$ coordinates are rigid at any order and from these we only perform a change of time. That is, we could write the metric in the system $S_{IG}$ in the form:
\begin{equation}\label{eq_34}
d{s^2} =  - d{\tau ^{*2}} + {\delta _{ij}}d{y^i}d{y^j} + O({y^2})
\end{equation}
understanding that we use $\left\{ {\lambda ,\vec y} \right\}$ coordinates and with ${\tau ^*}(\vec y,\lambda )$ being the potential.

Regarding the terms grouped under the symbol $O({y^2})$ (terms of order ${y^2}$ or higher), we observe that we could obtain a more accurate expression for the metric by solving   (\ref{eq_27}) for ${\tau ^*}(\vec z,\lambda )$, expanding  (\ref{eq_27}) in a power series around $\vec y = 0$ and using the condition   (\ref{eq_31}). As mentioned in the Introduction, these second-order terms would allow us to distinguish between FRC and FNC. We will not develop this line of work here because our current objective is to prove the feasibility of FRC and to do this it suffices to analyze the metric up to the first order. 

We can say that the $S_{IG}$ system supports rigid coordinates $\left\{ {\lambda ,\vec y} \right\}$ and in this sense is defined for all orders in $y$. $\lambda$ is the rigid time of the $S_{IG}$ system. In $S_{IG}$, we can define a new time $t$ and use the metric form  (\ref{eq_33}) or just use the metric form  (\ref{eq_34}).

If we take into account what we have seen so far, we can offer an implementation of the equivalence principle in the following terms:

\begin{quote}
\textit{If we have a space-time that, in a rigid reference system $S$, can be described by   (\ref{eq_20}), with a potential $\vec V$, then for each geodesic $G$ of the congruence $\vec V$ we can always define a rigid locally inertial reference system $S_{IG}$ such that its origin moves along the geodesic $G$ and rotates according to the vorticity of $\vec V$ evaluated at $G$. In the neighborhood of the origin of $S_{IG}$, it is possible to define a time with respect to which test particles move freely.}
\end{quote}

The extension of this result to any geodesic, unlike the Newtonian case, is not at all evident. In Section
$\S$\ref{s_6} we deal with the case of a general space-time. First, we will give a simple example to illustrate how things work.

\subsection{Fermi rigid coordinates for radial escape geodesics in Schwarzschild space-time}\label{s_5_1}
Schwarzschild space-time in a rigid reference system $S$ can be written in the Painlev\'e-Gullstrand  form  (\ref{eq_20}), with: 
\begin{equation}\label{eq_35}
\vec V = \sqrt {\frac{K}{{\left| {\vec x} \right|}}} \;\frac{{\vec x}}{{\left| {\vec x} \right|}}
\end{equation}
Under these coordinates, all radial geodesics $\vec x(\lambda )$ that are solutions of $\dot {\vec x} = \vec V$, that is to say the family of geodesics:
\begin{equation}\label{eq_36}
\vec x(\lambda ) = {\left[ {\frac{3}{2}\sqrt K \lambda  + {{\left| {{{\vec x}_0}} \right|}^{\frac{3}{2}}}} \right]^{\frac{2}{3}}}\frac{{{{\vec x}_0}}}{{\left| {{{\vec x}_0}} \right|}}
\end{equation}
has the same proper time $\lambda$. They are escape trajectories, i.e., outgoing trajectories with null speed at infinity. We note that if we want to study falling trajectories, we would need to consider Schwarzschild space-time with different rigid space-time coordinates. The \emph{good} expression for the metric would be the one we obtained with the potential $\vec V =  - \sqrt {\frac{K}{{\left| {\vec x} \right|}}} \;\frac{{\vec x}}{{\left| {\vec x} \right|}}$. In Section $\S$\ref{s_7_5}, we discuss this kind of problem in more detail.

We can find FRC for each of the geodesics $G$ in    (\ref{eq_36}). First, we make a rigid transformation $\vec x \to \vec y$ to give us the rigid locally inertial reference system $S_{IG}$, based on the congruence   (\ref{eq_35}). We do not need to make a rotation because the congruence  (\ref{eq_35}) is irrotational. We just need to perform the translation $\vec x = \vec y + \vec x(\lambda )$. We will then have:
\begin{equation}
\begin{array}{l}
d{s^2} =  - d{\lambda ^2} + {\left( {d\vec y - \vec V'd\lambda } \right)^2}\\
\vec V' = \sqrt {\frac{K}{{\left| {\vec y + \vec x(\lambda )} \right|}}} \;\frac{{\vec y + \vec x(\lambda )}}{{\left|
{\vec y + \vec x(\lambda )} \right|}} - \dot {\vec x}(\lambda )
\end{array}
\end{equation}
Now, we make a gauge transformation to eliminate the potential ${\vec V'}$ up to order $y^2$. According to   (\ref{eq_32}), this will define the time $t$ in $S_{IG}$:
\begin{equation}
t = \lambda  - \frac{{{{\vec y}^2}}}{4}\sqrt {\frac{K}{{{{\left| {\dot {\vec x}(\lambda )} \right|}^3}}}}  + O({y^2})
\end{equation}
We can verify that the $\left\{ {t,\vec y} \right\}$ coordinates are FC for the chosen geodesic   (\ref{eq_36}).

\section{General space-time}\label{s_6}
We will consider a general space-time which admits rigid covariant coordinates in the sense explained in  \cite{04-jaen_4}. The metric of such a space-time in a rigid reference system $S$ can be expressed as  \cite{04-jaen_4}:
\begin{equation}\label{eq_39}
d{s^2} =  - {\Phi ^2}d{\lambda ^2} + 2{K_i}\;d{x^i}\;d\lambda  + {\gamma _{ij}}d{x^i}d{x^j}
\end{equation}
with:
\begin{equation}\label{eq_40}
{\gamma _{ij}} = \frac{1}{{{{\cal H}^2}}}{\delta _{ij}} - {\sigma _{,{x^i}}}\;{\sigma _{,{x^j}}}
\end{equation}
The six gravitational potentials in this metric expression are $\left\{ {{\Phi},{K_i},{{\cal H}},\sigma } \right\}$.

If $U = {\partial _\lambda } + {V^i}\left( {\vec x,\lambda } \right){\partial _{{x^i}}}$ is a time-like geodesic congruence of the space-time  (\ref{eq_39}), with proper time $\tau \left( {\vec x,\lambda } \right)$, we can express ${\Phi}$ and $K_i$ in terms of the potentials $\tau$ and $\vec V$ in the form: 
\begin{equation}
\begin{array}{*{20}{l}}
{{\Phi ^2} =  - {\gamma _{ij}}{V^i}{V^j} + \tau _{,\lambda }^2 - {{\left( {{\tau _{,{x^i}}}{V^i}} \right)}^2}}\\
{{K_i}\; =  - {\gamma _{ij}}{V^j} - \left( {{\tau _{,\lambda }} + {\tau _{,{x^j}}}{V^j}} \right){\tau _{,{x^i}}}}
\end{array}
\end{equation}
The manifestly rigid covariant potentials of general relativity are $\left\{ {\tau ,\vec V,\;{\cal H},\sigma \;} \right\}$ and we can write  (\ref{eq_39}) in the manifestly  rigid covariant form as:
\begin{equation}\label{eq_42}
d{s^2} =  - d{\tau ^2} + \left( {{\tau _{,{x^i}}}{\tau _{,{x^j}}} + {\gamma _{ij}}} \right)(d{x^i} - {V^i}d\lambda
)(d{x^j} - {V^j}d\lambda )
\end{equation}
where $d\tau  = d\tau (\vec x,\lambda )$, even if we do not specify this.  (\ref{eq_39}) (or  (\ref{eq_42}) ) is shape invariant under the usual rigid transformations, such as those performed in the Newtonian case. The potential $\vec V$ transforms as a velocity ((\ref{eq_10}) remains valid) and the potentials $\tau ,{\cal H}$ and $\sigma$ as scalar fields.

The Lagrangian associated with   (\ref{eq_39}) is, if we adapt the factor so that $Ld\lambda $ becomes the proper time: 
\begin{equation}
L = \sqrt {{\Phi ^2} - 2{K_i}\;{{\dot {x}}^i}\; - {\gamma _{ij}}{{\dot {x}}^i}{{\dot {x}}^j}} 
\end{equation}
and the corresponding Hamiltonian is:
\begin{equation}
H(\vec x,\vec p,\lambda ) =  - {K_i}\;{\gamma ^{ij}}{p_j} - \sqrt {\left[ {1 + {\gamma ^{ij}}{p_i}{p_j}} \right]\left[
{{\gamma ^{ij}}{K_i}\;{K_j} + {\Phi ^2}} \right]} 
\end{equation}
where ${\gamma ^{ij}}$ is the inverse matrix of ${\gamma _{ij}}$, i.e., ${\gamma _{im}}{\gamma ^{mj}} = \delta _i^{\;j}$.

The metric   (\ref{eq_39}) can be constructed from any geodesic congruence $U$. That is,  (\ref{eq_39}) is invariant under gauge transformations $\left\{ {\vec V,\tau } \right\} \to \left\{ {{{\vec V}^*},{\tau ^*}} \right\}$ with $\tau^*$ being any solution of the Hamilton-Jacobi equation:
\begin{equation}\label{eq_45}
{\partial _\lambda }{\tau ^*} + H(\vec x,\vec p = {\partial _{\vec x}}{\tau ^*},\lambda ) = 0
\end{equation}
and ${\vec V^*}$ defined by: 
\begin{equation}\label{eq_46}
{\vec V^*} = \frac{{\partial H}}{{\partial \vec p}}(\vec x,\vec p = {\partial _{\vec x}}{\tau ^*},\lambda )
\end{equation}

\subsection{General rigid transformations}\label{s_6_1}
Given a metric of the form   (\ref{eq_39}), with  (\ref{eq_40}), we will say that a transformation of the space-time coordinates, $\left\{ {\lambda ,\vec x} \right\} \to \left\{ {\zeta ,\vec y} \right\}$, is a general rigid transformation if the transformed metric has the same rigid form as the original, but it may have a different rigid time $\zeta$. In Section $\S$4 of the reference  \cite{04-jaen_4}, we saw how to find rigid coordinates from a metric expressed in arbitrary space-time coordinates. If the starting metric in Section $\S$4 of reference  \cite{04-jaen_4} is already in rigid coordinates, the procedure described there will define a general rigid transformation. 

Given all the general rigid transformations, we distinguish the set that transform time, which we will call $\lambda$-rigid transformations, from the rest that do not transform time, $\zeta  = \lambda $. 

The $\lambda$-rigid transformations are characterized by the procedure mentioned above, in Section 4 of reference \cite{04-jaen_4}, when the starting metric, expression (\ref{eq_12}) of  \cite{04-jaen_4}, is already in rigid coordinates. That is, the coordinates $\{ {T,\vec X} \}$ of the metric (\ref{eq_12}) of  \cite{04-jaen_4} are rigid coordinates so ${C_{ij}} = {\delta _{ij}} - {\sigma _{,{X^i}}}\;{\sigma _{,{X^j}}}$ and the time $T$ is the starting rigid time. Non-trivial solutions for $T( {\lambda ,\vec X} )$ will define $\lambda$-rigid transformations. Once a solution has been found for $T( {\lambda ,\vec X} )$, it will be necessary to complete the work by searching for rigid Euclidean spatial coordinates $\vec x$ for the new time $\lambda$, as described in  \cite{04-jaen_4}. For any $\lambda$-rigid transformation $T( {\lambda ,\vec X} )$, there always exists a broad set of rigid Euclidean spatial coordinates $\vec x$. 

Within the second set, that of the transformations that do not change time, we find the subset of the usual rigid motions: rotations and translations. They are characterized by the fact that they leave invariant the element:
\begin{equation}
\bar d{s^2} = {\delta _{ij}}\bar d{x^i}\bar d{x^j}
\end{equation}
where $\bar d$ is the restriction of $d$ at the hypersurface $d\lambda=0$. To distinguish these transformations from those explained in the following section, we will call them $\delta$-rigid transformations.

\subsection{$\gamma$-rigid transformations }\label{s_6_2}
Another subset of the general rigid transformations, also belonging to the set that do not transform time, are those that leave invariant the element: 
\begin{equation}
\bar d{s^2}  = \left( {\frac{1}{{{{\cal H}^2}}}{\delta _{ij}} - {\sigma _{,{x^i}}}{\sigma _{,{x^j}}}} \right)\bar
d{x^i}\bar d{x^j}
\end{equation}
and are not $\delta$-rigid transformations.

We will restrict our study to the case in which ${\cal H} = 1$. The transformations we are looking for must leave invariant the element:
\begin{equation}
\bar d{\vec x^2} - \bar d{\sigma ^2} = {\eta _{\alpha \beta }}\bar d{z^\alpha }\bar d{z^\beta }
\end{equation}
where ${z^\alpha } = \left\{ {\sigma ,\vec x} \right\}$ and ${\eta _{\alpha \beta }} = {\rm{diag}}( - 1,1,1,1)$. This defines a kind of Poincar\'{e} space transformation that we can write as: 
\begin{equation}
{z'^\alpha } = L_\beta ^\alpha (\lambda )\left( {{z^\beta } - {Z^\beta }(\lambda )} \right)
\end{equation}
where $z{'^\alpha } = \left\{ {\sigma ',\vec y} \right\}$. The transformation is characterized by the functions $\{{{Z^\beta }(\lambda ),L_\beta ^\alpha (\lambda )} \}$ where $L_\beta ^\alpha (\lambda )$ is a Lorentz matrix: $L_\alpha ^\mu L_\beta ^\nu {\eta _{\mu \nu }} = {\eta _{\alpha \beta }}$. The transformation has the explicit appearance: 
\begin{equation}\label{eq_51}
\bar d{\vec x^2} - \bar d{\sigma ^2} = \bar d{\vec y^2} - \bar d\sigma {'^2}
\end{equation}
with:
\begin{equation}\label{eq_52}
\begin{array}{*{20}{l}}
{{y^i} = L_j^i(\lambda )\left( {{x^j} - {Z^j}(\lambda )} \right) + L_0^i(\lambda )\left( {\sigma (\vec x,\lambda ) -
{Z^0}(\lambda )} \right)}\\
{\sigma '(\vec y,\lambda ) = L_j^0(\lambda )\left( {{x^j} - {Z^j}(\lambda )} \right) + L_0^0(\lambda )\left( {\sigma
(\vec x,\lambda ) - {Z^0}(\lambda )} \right)}
\end{array}
\end{equation}
The first expression in (\ref{eq_52}) is the transformation of the coordinates we are looking for. The second expression is the transformation law of the potential $\sigma$ so that  (\ref{eq_51}) is satisfied. If, from  (\ref{eq_52}), we rule out $\delta$-rigid transformations (rotations and translations), then what we have left are the transformations that we can call $\gamma$-rigid transformations: 
\begin{equation}\label{eq_53}
\begin{array}{l}
\vec y = \vec x - \Upsilon \vec \beta \sigma  + \frac{{\Upsilon  - 1}}{{{\beta ^2}}}\left( {\vec \beta \cdot\vec x}
\right)\vec \beta \\
\sigma ' = \Upsilon \left( {\sigma  - \vec \beta \cdot\vec x} \right)
\end{array}
\end{equation}
where $\vec \beta $ is a function of $\lambda$ and $\Upsilon  \equiv \frac{1}{{\sqrt {1 - {\beta ^2}}}}$. Note that $\sigma (\vec x,\lambda )$, which is involved in the $\gamma$-rigid transformation, is set by the metric. 

Under an arbitrary transformation of space coordinates, $\vec x\; = \vec x(\vec y,\lambda )$, the metric  (\ref{eq_42}) becomes:
\begin{equation}
d{s^2} =  - d{\tau ^2} + \left( {{\tau _{,{y^i}}}{\tau _{,{y^j}}} + {{\gamma '}_{ij}}} \right)(d{y^i} - {V'^i}d\lambda
)(d{y^j} - {V'^j}d\lambda )
\end{equation}
where the potential ${{V'}^i}$ is:
\begin{equation}\label{eq_55}
{V'^i} = \frac{{\partial {y^i}}}{{\partial {x^m}}}\left( {{V^m} - \frac{{\partial {x^m}}}{{\partial \lambda }}} \right)
\end{equation}
The potential $\tau$ behaves like an scalar. If in addition the transformation is of the type   (\ref{eq_53}), then we will have:
\begin{equation}
{\gamma '_{ij}} = {\delta _{ij}} - {\sigma '_{,{y^i}}}{\sigma '_{,{y^j}}}
\end{equation}
where $\sigma ' = \frac{1}{{\sqrt {1 - {\beta ^2}} }}( {\sigma  - \vec \beta \cdot\vec x} )$. 

\subsection{Minkowski space-time}\label{s_6_3}
Now, we want to study the ordinary Poincar\'e transformations in the context of rigid general relativity. The decomposition of those transformations derived from this study is not a novelty and was already analyzed by M\"{o}ller \cite{19-moller} and used by Bel \cite{20-bel}.  

A standard observer $S$ in Minkowski space-time will use coordinates $\left\{ {\lambda ,\vec x} \right\}$:
\begin{equation}\label{eq_57}
d{s^2} =  - d{\lambda ^2} + d{\vec x^2}
\end{equation}
First, we perform a $\delta$-rigid transformation $\vec x = \vec y + \vec v\lambda $ with $\vec v = $constant. Using the $\left\{ {\lambda ,\vec y} \right\}$ coordinates, the metric becomes:
\begin{equation}\label{eq_58}
d{s^2} =  - d{\lambda ^2} + {\left( {d\vec y + \vec v\;d\lambda } \right)^2}
\end{equation}
that is to say, in $\left\{ {\lambda ,\vec y} \right\}$ coordinates, we have a non-null potential $\vec V =  - \vec v$. 

Second, from  (\ref{eq_58}), we perform a gauge transformation of the type $\{ {\tau  = \lambda ,\vec V =  - \vec v}\} \to \{ {{\tau ^*},{{\vec V}^*} = 0} \}$. The Lagrangian associated with  (\ref{eq_57}) is $L = \sqrt {1 - {{(\dot {\vec y} + \vec v)}^2}} $ and the Hamiltonian $H =  \vec p\cdot\vec v - \sqrt {1 + {p^2}} $. The equations that must be satisfied are: 
\begin{equation}
\begin{array}{*{20}{l}}
{{\partial _\lambda }{\tau ^*} + \vec v\cdot{\partial _{\vec y}}{\tau ^*} - \sqrt {1 + {{\left( {{\partial _{\vec
y}}{\tau ^*}} \right)}^2}}  = 0}\\
{0 = \vec v - \frac{{{\partial _{\vec y}}{\tau ^*}}}{{\sqrt {1 + {{\left( {{\partial _{\vec y}}{\tau ^*}} \right)}^2}}
}}}
\end{array}
\end{equation}
and the solution is:
\begin{equation}\label{eq_60}
{\tau ^*} = \sqrt {1 - {v^2}} \;\lambda  + \frac{{\vec v\cdot\vec y}}{{\sqrt {1 - {v^2}} \;}}
\end{equation}
In this gauge, the metric can be written using the potential ${{\tau ^*}}$: 
\begin{equation}\label{eq_61}
d{s^2} =  - d{\tau ^*}^2 + \left( {\tau _{,i}^*\tau _{,j}^* + {\delta _{ij}}} \right)d{y^i}d{y^j}
\end{equation}
This expression suggests performing a $\gamma$-rigid transformation, as described in Section $\S$\ref{s_6_2}, in such a way that we pass from $\sigma=0$ to $\sigma ' =  - \frac{{\vec v\cdot\vec y}}{{\sqrt {1 - {v^2}} \;}}$. Taking into account   (\ref{eq_53}), this is an $\vec y \to \vec z$ transformation with $\vec \beta  = \vec v$:
\begin{equation}\label{eq_62}
\vec z = \vec y + \frac{{\Upsilon  - 1}}{{{v^2}}}\left( {\vec v\cdot\vec y} \right)\vec v
\end{equation}
where $\Upsilon  \equiv \frac{1}{{\sqrt {1 - {v^2}} }}$. In this way, we cancel the term ${\tau_{,i}^*\tau _{,j}^*}$ from (\ref{eq_61}) with the new terms $ - {{\sigma '}_{,i}}{{\sigma '}_{,j}}$. In addition, being a transformation that does not depend on time $\lambda$, and since the original potential $\vec V$ is null, according to   (\ref{eq_55}), the transformed potential $\vec V'$ will also be null. That is to say, in $\left\{ {\lambda ,\vec z} \right\}$ coordinates, the metric is written:
\begin{equation}
d{s^2} =  - d{\tau ^*}^2 + d{{\vec z}^2}
\end{equation}
where ${\tau ^*}$ is now a function of $\left\{ {\lambda ,\vec z} \right\}$, ${\tau ^*}(\vec z,\lambda )$, according to   (\ref{eq_60}) and   (\ref{eq_62}). 

Third and finally, we perform the transformation $t = {\tau ^*}(\vec z,\lambda )$, thus obtaining the metric in Minkowski coordinates $\left\{ {t,\vec z} \right\}$. The composite transformation $\left\{ {\lambda ,\vec x} \right\} \to \left\{ {t,\vec z} \right\} $ is, of course, an ordinary Lorentz transformation. Thus, we see that we can understand a Lorentz transformation as a composition of general rigid transformations: a $\delta$-, a $\gamma$- and finally a $\lambda$-rigid transformation induced by a gauge transformation.

\subsection{$G$-isochronous geodesic congruence}\label{s_6_4}
Given a metric expressed in a rigid reference system $S$ with coordinates $\left\{ {\lambda ,\vec x} \right\}$, which we can write as   (\ref{eq_39}) with the potentials ${\Phi}$, $K_i$, $\cal {H}$ and $\sigma$ known, as functions of $\left\{ {\lambda ,\vec x} \right\}$, any time-like geodesic congruence $\{ {\tau\left( {\vec x,\lambda } \right),\vec V\left( {\vec x,\lambda } \right)} \}$ satisfies equations (\ref{eq_45}) and (\ref{eq_46}):
\begin{equation}\label{eq_64}
\left\{ {\begin{array}{*{20}{l}}
{{\partial _\lambda }\tau \left( {\vec x,\lambda } \right) + H\left( {\vec x,\vec p = {\partial _{\vec x}}\tau (\vec
x,\lambda ),\lambda } \right) = 0}\\
{\vec V\left( {\vec x,\lambda } \right) = \frac{{\partial H}}{{\partial \vec p}}\left( {\vec x,\vec p = {\partial _{\vec
x}}\tau (\vec x,\lambda ),\lambda } \right)}
\end{array}} \right.
\end{equation}
Given a time-like geodesic $G$, $\left\{ {\vec x(\lambda ),\tau (\lambda )} \right\}$, we will say that the geodesic congruence $\{ {\tau \left( {\vec x,\lambda } \right),\vec V\left( {\vec x,\lambda } \right)} \}$ is $G$-isochronous if $G$ belongs to the congruence, that is to say:
\begin{equation}
\begin{array}{*{20}{l}}
{\dot {\vec x}(\lambda ) = \vec V\left( {\vec x(\lambda ),\lambda } \right)\;}\\
{\tau (\lambda ) = \tau \left( {\vec x(\lambda ),\lambda } \right)}
\end{array}
\end{equation}
and in addition it fulfills:
\begin{equation}
{\left. {\bar d\tau } \right|_G} = 0
\end{equation}
which explicitly is ${\partial _{\vec x}}\tau(\vec x(\lambda ),\lambda ) = 0$. We will also add the simple condition $\tau (\vec x(\lambda ),\lambda ) = \lambda $ so that we can say that $\lambda$ is the
proper time of the geodesic $G$.

It should be noted that if $G$ is $\left\{ {\tau (\lambda ),\vec x(\lambda )} \right\}$, and the congruence $\{ {\tau \left( {\vec x,\lambda } \right),\vec V\left( {\vec x,\lambda } \right)} \}$ is $G$-isochronous, then the following will be satisfied: 
\begin{equation}\label{eq_67}
\dot {\vec x}(\lambda ) = \frac{{\partial H}}{{\partial \vec p}}(\vec x(\lambda ),\vec p = 0,\lambda )
\end{equation}
It is worth noting that, given the metric, this equation is determined by the choice of the time coordinate $\lambda$. Thus, once time $\lambda$ is set, only the geodesics $G$ that meet   (\ref{eq_67}) can belong to a $G$-isochronous congruence. In general, given a geodesic $G$, in order to find its $G$-isochronous congruence, it will be necessary to find a suitable time $\lambda$. It will be necessary to make $\lambda$-rigid transformations.

\section{Fermi rigid coordinates for a general space-time }\label{s_7}
Given a space-time and, in it, a specific time-like geodesic, we want to implement
the equivalence principle by finding rigid coordinates that are, in turn, FC for the given geodesic.

\subsection{First step. Fixing the rigid time: $\lambda$-rigid transformation}\label{s_7_1}
We consider an arbitrary space-time and in it, an arbitrary time-like geodesic $G$. The first and most difficult step is to set the rigid time coordinate $\lambda$ for which there exists a $G$-isochronous geodesic congruence. It may be the case that:
\begin{enumerate}

 \item From the beginning of the process we have a rigid system of coordinates. We will check whether the geodesic $G$ satisfies the corresponding equation (\ref{eq_67}). If it does, then it will be possible to find the $G$-isochronous congruence by solving   (\ref{eq_64}). If equation   (\ref{eq_67}) is not satisfied, then it will be necessary to perform $\lambda$-rigid transformations.
 
\item We do not have a rigid coordinate system to express the metrics of our space-time. In this case, we will need to solve the problem to find a set of rigid coordinates, as described in Section $\S$4 of the reference \cite{04-jaen_4}, but restricted to the condition that the geodesic congruence potential must be $G$-isochronous. 
 
\end{enumerate}
We have no evidence that it is always possible to meet the above requirement, nor do we have examples to the contrary. However, the final purpose of this work is to construct a sufficiently broad theoretical framework whereby Fermi and rigid coordinates can be related for cases such as linear gravitational waves. We will work on this topic successfully in Section $\S$\ref{s_8}

In what follows, we will assume that we are able to write the space-time in a system of coordinates $\left\{ {\lambda,\vec x} \right\}$ in such a way that the geodesic $G$ is $\vec x(\lambda )$ and the potentials $\left\{ {\tau \left( {\vec x,\lambda } \right),\vec V\left( {\vec x,\lambda } \right)}\right\}$ are a $G$-isochronous congruence. This means that $\tau (\lambda ) = \tau(\lambda ,\vec x(\lambda )) = \lambda $, ${\partial _{\vec x}}\tau (\lambda ,\vec x(\lambda )) = 0$ and $\dot {\vec x}(\lambda ) = \vec V(\vec x(\lambda ),\lambda ) = \frac{{\partial H}}{{\partial \vec p}}(\vec x(\lambda ),\vec p = 0,\lambda )$. 

As a consequence, the metric expressed in the starting rigid system $S$ can be written in the form:
\begin{equation}\label{eq_68}
d{s^2} =  - d{\tau ^2} + {\gamma _{ij}}(d{x^i} - {V^i}d\lambda )(d{x^j} - {V^j}d\lambda ) + O\left( {{{\left| {\vec x -
\vec x(\lambda )} \right|}^2}} \right)
\end{equation}
with ${\gamma _{ij}} = {\delta _{ij}} - {\sigma _{,{x^i}}}\;{\sigma _{,{x^j}}}$. That is to say, in linear order, we can eliminate the terms ${\tau _{,{x^i}}}{\tau _{,{x^j}}}$. In this expression, there may be terms $O\left( {{{\left| {\vec x - \vec x(\lambda )} \right|}^2}} \right)$ in both ${V^i}$ and $\tau$. 

With this first step, undoubtedly the most difficult, we have defined a suitable rigid time $\lambda$. Consequently, the rigid motions and $\gamma$-rigid transformations will be defined using this time.

\subsection{Second step. $\gamma$-rigid transformation}\label{s_7_2}
From a metric of the form   (\ref{eq_68}) we make a $\gamma$-rigid transformation, $\vec x \to \vec y$, with the condition $\sigma ' = f(\lambda ) + O( {{{\left| {\vec x - \vec x(\lambda )} \right|}^2}} )$. This is achieved by a transformation such as   (\ref{eq_53}) and choosing:
\begin{equation}\label{eq_69}
\vec \beta (\lambda ) = {\partial _{\vec x}}\sigma \left( {\vec x(\lambda ),\lambda } \right)
\end{equation}
It is important to note that this transformation, although it does not leave the Euclidean metric unchanged, meets $O({{{\left| {\vec x - \vec x(\lambda )} \right|}^2}} ) = O( {{{\left| {\vec y - \vec y(\lambda )} \right|}^2}} )$ where $\vec y(\lambda )$ is the geodesic $\vec x(\lambda )$ in the new coordinates.

The result of this transformation on the metric  (\ref{eq_53}), will be a metric of the form: 
\begin{equation}\label{eq_70}
d{s^2} =  - d{\tau ^2} + {\left( {d\vec y - \vec V'd\lambda } \right)^2} + O\left( {{{\left| {\vec y - \vec y(\lambda )}
\right|}^2}} \right)
\end{equation}
where $\vec V'$ is defined by  (\ref{eq_55}), according to the transformation $\vec x \to \vec y$   (\ref{eq_53}) with  (\ref{eq_69}).

\subsection{Third step. $\delta$-rigid transformation}\label{s_7_3}
Starting from  (\ref{eq_70}), we perform a $\delta$-rigid transformation (ordinary translation and rotation) $\vec y \to \vec z$ defined according to the geodesic $\vec y(\lambda )$ and the congruence $\vec V'$. That is, defined as in the Newtonian case but now using the congruence $\vec V'$ derived from performing the $\gamma$-rigid transformation above. That is, given by:
\begin{equation}
\vec R(\lambda ) = \vec y(\lambda )\;\;;\;\;\;\vec \Omega (\lambda ) = \frac{1}{2}\left[ {{\partial _{\vec y}} \times
\vec V'} \right](\vec R(\lambda ),\lambda )
\end{equation}
In $\vec z$ coordinates the geodesic will be simply $\vec z = 0$ and $O\left( {{{\left| {\vec y - \vec y(\lambda )}\right|}^n}} \right) = O\left( {{z^n}} \right)$. We will arrive at the metric form:
\begin{equation}\label{eq_72}
d{s^2} =  - d{\tau ^2} + {\left( {d\vec z - \vec V''d\lambda } \right)^2} + O\left( {{z^2}} \right)
\end{equation}
where, as in the non-relativistic case:
\begin{equation}\label{eq_73}
\vec V''\left( {\vec z,\lambda } \right) = {\left[ {\left( {\vec z\cdot{\partial _{\vec y}}} \right)\vec V' +
\frac{1}{2}\vec z \times \left( {{\partial _{\vec y}} \times \vec V'} \right)} \right]_{\vec y = \vec R}} + O\left(
{{z^2}} \right)
\end{equation}
which does not include zero order but $z$ order. As in the Newtonian case, the property ${\partial _{\vec z}}\times \vec V'' = O({z^2})$ is met.

\subsection{Fourth step. The proper time coordinate $t$}\label{s_7_4}
As the congruence is $G$-isochronous, the potential $\tau $ meets ${\left. {\bar d\tau }
\right|_G} = 0$ which is now ${\partial _{\vec z}}\tau \left( {0,\lambda } \right) = 0$. That is to say, until now we have a known $\tau \left( {\vec z,\lambda } \right)$ that has the form $\tau \left( {\vec z,\lambda } \right) = \lambda  + O({z^2})$. We can still make $O({z^2})$ gauge transformations. This fact will allow us, starting from the metric  (\ref{eq_72}), to perform a gauge transformation:
\begin{equation*} 
\left\{ {\vec V'' = O\left( z \right),\tau  = O\left( {{z^2}} \right)} \right\} \to {\rm{ }}\left\{ {{\vec V''^*} = O\left( {{z^2}} \right),{\tau ^*} = O\left( {{z^2}} \right)} \right\}{\rm{ }}
\end{equation*}

The equations that must be satisfied are:
\begin{equation}\label{eq_74}
{\partial _\lambda }{\tau ^*} + H\left( {\vec z,\vec p = {\partial _{\vec z}}{\tau ^*},\lambda } \right) = 0
\end{equation}
\begin{equation}\label{eq_75}
{\vec V''^*} = \frac{{\partial H}}{{\partial \vec p}}\left( {\vec z,\vec p = {\partial _{\vec z}}{\tau ^*},\lambda }
\right) = O\left( {{z^2}} \right)
\end{equation}

We will take into account that the geodesic congruence is $G$-isochronous, ${\partial _{\vec z}}\tau\left( {\vec z = 0,\lambda } \right) = 0$, and we also require ${\partial _{\vec z}}{\tau^*}\left( {\vec z = 0,\lambda } \right) = 0$. Up to order $O({z^2})$  (\ref{eq_74}) and  (\ref{eq_75}) become:
\begin{equation}
\left\{ {\begin{array}{*{20}{l}}
{{\partial _\lambda }{\tau ^*} = 1 + O({z^2})}\\
{{\partial _{\vec z}}{\tau ^*} = {\partial _{\vec z}}\tau \;\; + \vec V'' + O({z^2})}
\end{array}} \right.
\end{equation}
We note that the integrability conditions are again guaranteed because ${\partial _{\vec z}} \times \vec V'' = O({z^2})$ and by the fact that we are neglecting terms of order $O({z^2})$. As can be seen by direct calculation (explicitly using expression (\ref{eq_73}) for $\vec V''$), the solution is:
\begin{equation}\label{eq_77}
{\tau ^*}\left( {\vec z,\lambda } \right) = \tau \left( {\vec z,\lambda } \right)\;\; + \frac{1}{2}\vec z\cdot\vec V'' +
O\left( {{z^3}} \right)
\end{equation}
Using the space-time coordinates $\left\{ {\lambda ,\vec z} \right\}$ and this gauge, the metric can be written as:
\begin{equation*}
d{s^2} =  - d{\tau ^{*2}} + d{\vec z^2} + O\left( {{z^2}} \right)
\end{equation*}
which means that these rigid coordinates define the rigid locally inertial reference system attached to the geodesic $G$: $S_{IG}$.

We define the time coordinate $t$ of the system $S_{IG}$ according to $t = {\tau ^*}\left( {\vec z,\lambda} \right)$. Making the expression explicit, the result is:
\begin{equation}\label{eq_78}
t = \tau \;(\vec z,\lambda ) + \frac{1}{2}\vec z\cdot{\left[ {\left( {\vec z\cdot{\partial _{\vec y}}} \right)\vec V' +
\frac{1}{2}\vec z \times \left( {{\partial _{\vec y}} \times \vec V'} \right)} \right]_{\vec y = \vec R}} + O\left(
{{{\left| {\vec z} \right|}^3}} \right)
\end{equation}
We should recall that in $\vec z$ coordinates, the geodesic is $\vec z = 0$ and the congruence $G$-isochronous, i.e., ${\partial _{\vec z}}\tau \left( {\vec z = 0,\lambda } \right) = 0$.

In $\left\{ {t,\vec z} \right\}$ coordinates the metric takes the form: 
\begin{equation}
d{s^2} =  - d{t^2} + d{\vec z^2} + O\left( {{z^2}} \right)
\end{equation}
which means that the $\left\{ {t,\vec z} \right\}$ coordinates are FRC.

Regarding the terms grouped under the symbol $O\left( {{z^2}} \right)$ (terms of order $z^2$ or higher), we observe that we could obtain a more accurate expression for the metric by solving  (\ref{eq_74}) for ${\tau ^*}(\vec z,\lambda )$ with the condition  (\ref{eq_77}). The same comments that are made above after equation (\ref{eq_34}) apply here. 

In fact, beyond second order (first order in the metric), time $t$ and the metric no longer have clear physical meaning, and it may be preferable to work with the rigid time $\lambda$. In particular the transformation  (\ref{eq_78}) is a $\lambda$-rigid transformation as long as we remain in second order. What we can say, at any order, is that the $S_{IG}$ system supports rigid coordinates $\left\{{\lambda ,\vec z} \right\}$ and in this sense $S_{IG}$ is defined for all orders in $z$. 

As far as the principle of equivalence is concerned, we can conjecture the following statement, which we call the rigid equivalence principle.
\begin{quote}
\textit{\textbf{Rigid equivalence principle}: in any space-time and for any time-like geodesic $G$, it is always possible to construct a locally inertial rigid reference system $S_{IG}$ whose origin moves with the geodesic G and whose rotation coincides with the vorticity of a $G$-isochronous geodesic congruence at the origin. In the neighborhood of the origin of $S_{IG}$, it is possible to define a time with respect to which test particles move freely.}
\end{quote}

\subsection{Fermi rigid coordinates for radial time-like geodesics in Schwarzschild space-time}\label{s_7_5}
We can apply what we have learned and find $S_{IG}$, thus finding FRC, for a Schwarzschild radial time-like geodesic $G$, beyond the escape geodesics. In order not to lengthen the study, we will only detail the first step, which is the only one that is not automatic. Once this first step has been carried out, the others can be followed without any more explanation than that given in the previous section, which we only mention here. 

To cover the first step, we begin with the following stationary rigid covariant spherical symmetry form of the metric:
\begin{equation}
d{s^2} =  - d{\lambda ^2} + \left( {1 - \sigma _{,r}^2} \right){\left( {d{r^2} - Vd\lambda } \right)^2} + {r^2}d{\Omega
^2}
\end{equation}
where $d{\Omega ^2} = d{\theta ^2} + {\sin ^2}\theta \;d{\phi ^2}$, the potentials $\sigma$ and $V$ are functions of the radial coordinate $r$ and the potential $\tau \;(r,\lambda )$ is simply $\tau \;(r,\lambda ) = \lambda $. Solving Einstein's field equations in empty space we find:
\begin{equation}
V =\pm \sqrt {\frac{K}{r} + V_0^2} \;\;\;;\;\;\;\sigma  = \;\sqrt {\frac{{V_0^2}}{{1 + V_0^2}}} \;r
\end{equation}
If $K=0 $ then $V=V_0$ and the space-time is Minkowski, written using a constant velocity $V_0$ radial congruence. If $V_0=0$ we recover the usual Painlev\'{e}-Gullstrand form for the Schwarzschild metric. However, the present form is not simply a rigid or gauge transformation from $V_0=0$ to try to include the required geodesics. These transformations would not lead us to the $G$-isochronous condition ${\left. {\bar d\tau } \right|_G} = 0$ for radial non-escape geodesics. What is important here is that we have found a solution that meets the requirements for implementing the rest of the steps outlined in the previous section, ${\left. {\bar d\tau } \right|_G} = 0$ because in fact $\tau  = \lambda $. We have found a rigid covariant form for the Schwarzschild metric with a suitable time coordinate $\lambda$ for the required geodesic radial non-escape geodesic. For each family of radial geodesics characterized by a different value of $V_0$ and a different sign $\pm$ of $V$, we have a different coordinate system with a different time, $\lambda$. These systems are related by $\lambda$-rigid transformations.

By setting a value of $V_0$ and a sign of $V$, say $+$, and using Euclidean space coordinates, we have:
\begin{equation}
\vec V = \sqrt {\frac{K}{{\left| {\vec x} \right|}} + V_0^2} \;\frac{{\vec x}}{{\left| {\vec x} \right|}}\;;\;\;\;\sigma
 = \;\sqrt {\frac{{V_0^2}}{{1 + V_0^2}}} \;\left| {\vec x} \right|\;
\end{equation}
The second step will be to perform a $\gamma$-rigid transformation, $\vec x \to \vec y$, with $\vec \beta (\lambda ) = {\partial _{\vec x}}\sigma \left( {\vec x(\lambda ),\lambda } \right)$, where $\vec x(\lambda )$ is the chosen geodesic $G$ solution of $\dot {\vec x}(\lambda ) = \vec V\left( {\vec x(\lambda )} \right)$. The new potential $\vec V'$ will be given according to (\ref{eq_55}). 

The third step will consist of a $\delta$-rigid transformation $\vec y \to \vec z$ towards the locally inertial reference system $S_{IG}$ based on the $\vec V'$ potential that emerged in the previous step. The new potential will be $\vec V''$.

The fourth step will consist of eliminating the first order in $z$ of $\vec V''$ by a gauge transformation with $ {\tau ^*}\left( {\vec z,\lambda } \right)$ defined by (\ref{eq_77}), and finally defining time $t$ according to $t = {\tau ^*}\left( {\vec z,\lambda } \right)$ or   (\ref{eq_78}).

\section{Fermi rigid coordinates for linear plane gravitational waves}\label{s_8}
In this section, we apply what we have learned in order to find FRC for a linear plane gravitational wave or, in other words, to find a rigid locally inertial reference system for these waves. 

In the reference  \cite{04-jaen_4} we find a set of rigid coordinates $\left\{ {\lambda ,\vec x}\right\}$ for a linear plane gravitational wave originally written in Gaussian coordinates $\{ {T,\vec X} \}$ as:
\begin{equation}\label{eq_83}
d{s^2} =  - d{T^2} + d{\vec X^2} + {\varepsilon ^2}\left\{ {2{h_ \times }dX\;dY + {h_ + }(d{X^2} - d{Y^2})} \right\}
\end{equation}
The $\{ {T,\vec X} \}$ coordinates are Gaussian so that the points at rest $\vec X = {\vec X_0}$ form a family of time-like geodesics of the proper time $T$. If the wave ``disappears'', i.e. if ${h_\times } = {h_ + } = 0$, the geodesic $\vec X = {\vec X_0}$ will consist of points at rest in Minkowski space-time with standard coordinates. This is why we are interested in these geodesics.

We want to find FRC for one of the geodesics, say $G$, of this family. We can always adjust the coordinates so that the required geodesic becomes $\vec X = 0$. 

In this section, and as far as wave linearity is concerned, we will work up to order $\varepsilon ^2$, even if we do not explicitly indicate it.

\subsection{First step}\label{s_8_1}
We have to find rigid coordinates for   (\ref{eq_83}) such that the geodesic congruence is $G$-isochronous. Part of this work was already performed in reference  \cite{04-jaen_4} for the same kind of waves. Equations (17) of  \cite{04-jaen_4} were solved with $\tau (\vec X,\lambda ) = T(\vec X,\lambda ) = \lambda  + \varepsilon (X{T_{(x)}} + Y{T_{(y)}})$, $\sigma (\vec X,\lambda ) = \varepsilon(X{\sigma _{(x)}} + Y{\sigma _{(y)}})$ where $\{T_{(x)},T_{(y)},\sigma _{(x)},\sigma _{(y)}\}$ are functions of $Z-\lambda$, and the equations that must be satisfied are:\footnote{In this section we will use a prime, $f' $, to indicate the derivation of a function with respect to its argument.}

\begin{equation}\label{eq_84}
\begin{array}{l}
 {\sigma_{(x)}}'^2  + {\sigma_{(y)}}'^2  = {T_{(x)}}'^2  + {T_{(y)}}'^2  \\ 
 2{\sigma _{(x)}}'{\sigma_{(y)}}' - 2{T_{(x)}}' {T_{(y)}}' + {h_\times}''= 0 \\ 
 2{\sigma_{(x)}}'^2  - 2 {T_{(x)}}'^2  + {h_+}'' = 0 \\ 
 2 {T_{(y)}}'^2  - 2{\sigma_{(y)}}'^2  + {h_+}'' = 0 \\ 
 \end{array}
\end{equation}
If these conditions are met, we arrive at a metric form for   (\ref{eq_83}), given by (\ref{eq_39}) with (\ref{eq_40}), but instead of having $\delta_{ij}$ in   (\ref{eq_40}) we have $\Delta_{ij}$, which is also a flat space metric but not in the Euclidean form. Because of this, we have to solve $\Delta _{ij}dX^{i}dX^{j}=\delta_{ij}dx^{i}dx^{j}$ for a coordinate change $\vec{X}=\{X,Y,Z\}\to \vec{x}=\{x,y,z\}$. This change depends on $\lambda$, which in the space $\vec{X}$ acts as a parameter. This work was also completed in the reference  \cite{04-jaen_4}. Linking the two transformations, $\{X,Y,Z,T\}\to \{x,y,z,\lambda\}$ and up to order $\varepsilon ^{2}$, we have the explicit result for the cross mode ($h_+=0$):
\begin{equation}\label{eq_85}
\begin{array}{l}
X = x + {\varepsilon ^2}y\left[ {{T_{(x)}}{T_{(y)}} - \frac{{{h_ \times }}}{2} + I + f(\lambda )} \right]\\
Y = y + {\varepsilon ^2}x\left[ {{T_{(x)}}{T_{(y)}} - \frac{{{h_ \times }}}{2} - I - f(\lambda )} \right]\\
Z = z + \frac{{{\varepsilon ^2}}}{2}xy{h_ \times }^\prime \\
T = \lambda  + \varepsilon \left( {x{T_{(x)}} + y{T_{(y)}}} \right)
\end{array}
\end{equation}
where $I \equiv {\rm{ }}\int {\left( {{T_{(x)}}{T_{(y)}}^\prime  - {T_{(y)}}{T_{(x)}}^\prime } \right)dz} $ will be understood as a primitive, without an integration constant, because we have already taken it into account through the arbitrary function ${f(\lambda )}$. We want to clarify that in equation (27) of reference  \cite{04-jaen_4}, we made a small error in considering that each integral of (24)-\cite{04-jaen_4} provided an independent function when writing (27)-\cite{04-jaen_4}. Only if ${f_x}(\lambda ) =  - {f_y}(\lambda ) = f(\lambda )$ (27)-\cite{04-jaen_4} is there a change towards rigid coordinates. It should be noted that, in \cite{04-jaen_4}, the subsequent election of these functions fulfilled the condition ${f_x}(\lambda ) =  - {f_y}(\lambda )$. This is why here, in (\ref{eq_85}), we express (24)-\cite{04-jaen_4} making the arbitrary functions explicit. 

Unlike  \cite{04-jaen_4}, where the objective was to find some rigid coordinate system and therefore the arbitrary functions of $\lambda$ were fixed without a specific criterion, we now impose conditions with physical meaning that allow us to fix these functions. The main condition is to make the congruence $\vec X = {\vec X_0}$ $G$-isochronous, with $G$ being the  geodesic selected: $\vec X = 0$. Because the proper time of the congruence is $T$ and the geodesic $G$ is $\vec x =0$, in the rigid coordinate system $\left\{ {\lambda ,\vec x} \right\}$ the $G$-isochronous condition is:
\begin{equation}
{\partial _{\vec x}}T(\vec x = 0,\lambda ) = 0
\end{equation}

There is a secondary condition which, if not imposed, we could overcome with a $\gamma$-rigid transformation when we are carrying out the second step: the change (\ref{eq_85}) in the metric  (\ref{eq_83}) will define the rigid potentials and in particular the potential $\sigma$. Of that potential, we require that: 
\begin{equation}\label{eq_87}
{\partial _{\vec x}}\sigma (\vec x = 0,\lambda ) = 0
\end{equation}
Taking into account these conditions, together with   (\ref{eq_84}), we obtain the conditions on the functions ${T_{(x)}},{T_{(y)}}$ and $f(\lambda)$: 
\begin{equation}\label{eq_88}
{T_{(x)}}\left( {0 - \lambda } \right) = 0\;;\;{T_{(y)}}\left( {0 - \lambda } \right) = 0\;;f(\lambda ) = 0
\end{equation}
where $0-\lambda$ means that we have taken $z=0$. 

The fulfillment of these conditions ensures that  (\ref{eq_85}) is a change towards rigid coordinates with the geodesic congruence $\vec X = {\vec X_0}$ being $G$-isochronous, where $G$ is the geodesic $\vec X = 0$ and  also satisfies  (\ref{eq_87}).

A particularly interesting case is that of the cross mode of a monochromatic linear plane wave, with frequency $\omega$. This corresponds to considering (\ref{eq_83}) with $h_+=0$ and:
\begin{equation}
\varepsilon ^{2}\;h_{\times}(Z,T)=A^{2}\sin (\omega (Z-T))
\end{equation}
i.e., $\varepsilon =A$ and $h_{\times}=\sin [\omega (Z-\lambda )]$. 

As a solutions of   (\ref{eq_84}), also satisfying   (\ref{eq_88}), we choose:
\begin{equation}
\begin{array}{l}
{T_{(x)}} = \sqrt 2 \sin [\frac{\omega }{2}(Z - \lambda )]\; + \sqrt 2 \sin [\frac{\omega }{2}\lambda ]\;\\
{T_{(y)}} = \sqrt 2 \cos [\frac{\omega }{2}(Z - \lambda )] - \sqrt 2 \cos [\frac{\omega }{2}\lambda ]
\end{array}
\end{equation}
From   (\ref{eq_85}) and recalling that, according to (\ref{eq_88}), we set $f(\lambda ) = 0$: 
\begin{equation}\label{eq_91}
\begin{array}{*{20}{l}}
{X = x + {A^2}\frac{1}{2}y\left\{ {\sin \left[ {\omega (z - \lambda )} \right] + 4\sin \left[ {\frac{\omega }{2}z}
\right] - 4\sin \left[ {\omega (\frac{z}{2} - \lambda )} \right] - 2\sin \left[ {\omega \lambda } \right] - 2\omega z}
\right\}}\\
{Y = y + {A^2}\frac{1}{2}x\left\{ {\sin \left[ {\omega (z - \lambda )} \right] + 4\sin \left[ {\frac{\omega }{2}z}
\right] - 4\sin \left[ {\omega (\frac{z}{2} - \lambda )} \right] - 2\sin \left[ {\omega \lambda } \right] - 2\omega z}
\right\}}\\
{Z = z + {A^2}\frac{1}{2}\omega xy\cos \left[ {\omega (z - \lambda )} \right]}\\
{T = \lambda  + \sqrt 2 A\left\{ {x\left( {\sin \left[ {\frac{\omega }{2}(z - \lambda )} \right] + \sin \left[
{\frac{\omega }{2}\lambda } \right]} \right) + y\left( {\cos \left[ {\frac{\omega }{2}(z - \lambda )} \right] - \cos
\left[ {\frac{\omega }{2}\lambda } \right]} \right)} \right\}}
\end{array}
\end{equation}

\subsection{Second step. $\gamma$-rigid transformation}\label{s_8_2}
This step will not be necessary as we have been able to set ${\partial _{\vec x}}\sigma (\vec x = 0,\lambda ) =
0$ during the first step. However, simply for consistency of notation we will set: $\vec y = \vec x$.

\subsection{Third step. $\delta$-rigid transformation}\label{s_8_3}
From   (\ref{eq_91})  we can calculate the potential $\vec V$ associated with the geodesics $\vec X =$ constant. We obtain:
\begin{equation}\label{eq_92}
\begin{array}{*{20}{l}}
{\vec V = {A^2}y\left\{ {\frac{1}{2}\cos \left[ {\omega (z - \lambda )} \right] - 2\cos \left[ {\omega (\frac{z}{2} -
\lambda )} \right] + \omega \cos \left[ {\omega \lambda } \right]} \right\}{\partial _x} + }\\
{{A^2}x\left\{ {\frac{1}{2}\cos \left[ {\omega (z - \lambda )} \right] - 2\cos \left[ {\omega (\frac{z}{2} - \lambda )}
\right] + \omega \cos \left[ {\omega \lambda } \right]} \right\}{\partial _y} - }\\
{{A^2}\frac{1}{2}{\omega ^2}xy\sin \left[ {\omega (z - \lambda )} \right]{\partial _z}}
\end{array}
\end{equation}
which together with $\tau (\vec x,\lambda ) = T(\vec x,\lambda )$ defines the geodesic congruence as $G$-isochronous. 

From   (\ref{eq_91}), $\vec X=0$ implies $\vec x=0$, so it is not necessary to make a rigid translation: $\vec R(\lambda ) = 0$. Direct calculation of (\ref{eq_92}) shows us that $\vec \Omega(\lambda ) = \frac{1}{2}{\left[ {{\partial _{\vec x}} \times \vec V} \right]_{\vec x = 0}} = 0$ so that we will not need to make a rigid rotation either. Once again, only for consistency of notation we will set: $\vec z = \vec y = \vec x$.

The difference between the change we have now, expression (\ref{eq_91}), and that found in (28)-\cite{04-jaen_4} is that the latter leads us to possible rigid coordinates. Now, from all possible rigid coordinates, we select those in which the metric takes the form   (\ref{eq_72}), with a potential $\vec V'' = \vec V$, given by   (\ref{eq_92}), that fulfills the conditions of being null and irrotational on the geodesic. It still contains linear terms in the space coordinates. 

We can say that we have now found the rigid locally inertial reference system attached to the geodesic $\vec X=0$, $S_{IG}$, with coordinates $\left\{ {\lambda ,\vec x} \right\}$. The change   (\ref{eq_91})  globally defines $S_{IG}$, i.e., it is not limited to any order in $x$.

\subsection{Fourth step. The proper time coordinate $t$}\label{s_8_4}
In the final step, we perform the gauge transformation $\{ \vec V,\tau \}  \to \{ {{\vec V}^*} = O\left( {{x^2}} \right),{\tau ^*}\}$, defined in (\ref{eq_74})  (\ref{eq_75}), and define the time coordinate $t$ of $S_{IG}$ as $t = {\tau^*}\left( {\vec z,\lambda } \right)$. 

Taking into account   (\ref{eq_78}), recalling that $\vec y = \vec x$ and maintaining, for ease of calculation, $\vec z \ne \vec x$, we can write:
\begin{equation}
t = {\left\{ {T(\vec z,\lambda ) + \frac{1}{2}\vec z\cdot{{\left[ {\left( {\vec z\cdot{\partial _{\vec x}}} \right)\vec
V} \right]}_{\vec x = 0}}} \right\}_{\vec z = \vec x}} + O\left( {{{\left| {\vec x} \right|}^3}} \right)
\end{equation}
where, only after the derivation and the substitution of the geodesic $\vec x = 0$, do we make the substitution $\vec z = \vec x$. The result of this operation, taking into account expression  (\ref{eq_92}) for ${\vec V}$, is:
\begin{equation}
t = T(\vec x,\lambda ) - \frac{1}{2}{A^2}\omega xy\cos \left[ {\omega \lambda } \right] + O\left( {{{\left| {\vec x}
\right|}^3}} \right)
\end{equation}
With this information, we obtain the change from the original $\{ {T,\vec X} \}$ coordinates to FRC attached to the geodesic $\vec X=0$, $\{t,\vec x\} $:
\begin{equation}\label{eq_95}
\begin{array}{*{20}{l}}
{X = x + {A^2}\frac{1}{2}y\left\{ {\sin \left[ {\omega (z - t)} \right] + 4\sin \left[ {\frac{\omega }{2}z} \right] -
4\sin \left[ {\omega (\frac{z}{2} - t)} \right] - 2\sin \left[ {\omega t} \right] - 2\omega z} \right\}}\\
{Y = y + {A^2}\frac{1}{2}x\left\{ {\sin \left[ {\omega (z - t)} \right] + 4\sin \left[ {\frac{\omega }{2}z} \right] -
4\sin \left[ {\omega (\frac{z}{2} - \lambda )} \right] - 2\sin \left[ {\omega t} \right] - 2\omega z} \right\}}\\
{Z = z + {A^2}\frac{1}{2}\omega xy\cos \left[ {\omega (z - t)} \right]}\\
{T = t + \frac{1}{2}{A^2}\omega xy\cos \left[ {\omega t} \right] + O\left( {{{\left| {\vec x} \right|}^3}} \right)}
\end{array}
\end{equation}
Finally, we can now study the geodesics $\vec X = {\vec X_0}$. We can find the relationship between $\left\{ {{X_0},{Y_0},{Z_0}} \right\}$ and $\left\{ {t = 0,{x_0},{y_0},{z_0}} \right\}$ by making $t=0$ in expression (\ref{eq_95}). For the geodesics $\left\{ {t = 0,{x_0},{y_0},{z_0} = 0} \right\}$ we find the expressions: 
\begin{equation}
\begin{array}{*{20}{l}}
{x = {x_0} - {A^2}\frac{1}{2}{y_0}\sin \left[ {\omega \;t} \right]}\\
{y = {y_0} - {A^2}\frac{1}{2}{x_0}\sin \left[ {\omega t} \right]}\\
{z = {A^2}\frac{1}{2}\omega {x_0}{y_0}\left( {1 - \cos \left[ {\omega \;t} \right]} \right)}\\
{T = t + \frac{1}{2}{A^2}\omega {x_0}{y_0}\cos \left[ {\omega t} \right] + O\left( {{{\left| {{{\vec x}_0}} \right|}^3}}
\right)}
\end{array}
\end{equation}
Up to order $O\left( {\left| {{{\vec x}_0}} \right|} \right)$, this coincides with the usual result. Of course, up to order $O\left( {\left| {{{\vec x}_0}} \right|} \right)$, the metric will have the standard Minkowski form. This result is also valid up to order $O( {{{\left| {{{\vec x}_0}} \right|}^2}} )$. In contrast, up to order $O( {{{\left| {{{\vec x}_0}} \right|}^2}} )$, the metric will no longer has a Minkowski form, although it will still have a rigid form.

\section{Conclusions}
In this paper, we have progressed further and moved deeper into the previous study  \cite{04-jaen_4}, where we found a set of rigid coordinate systems for linear plane gravitational waves. Two related issues remained unanswered: how to arrive at a rigid coordinate system from among the set found; and what relation those systems have with the commonly used Fermi coordinates.

Our goal was to find coordinates that were both rigid and Fermi. To do this, we have analyzed rigid covariant transformations beyond the usual rigid motions.

Given any space-time and, in it, any time-like geodesic, we have identified four steps that allow us to construct a rigid coordinate system that is also a Fermi coordinate system. We call them: Fermi rigid coordinates (FRC), as opposed to Fermi normal coordinates (FNC).

The first step consists of representing the space-time in rigid coordinates that fulfill the condition of being $G$-isochronous for the chosen geodesic. This step is important because it provides us with a time coordinate with which to define generalized rigid transformations and thereby to arrive at the locally inertial system. Although we have been able to apply this step to several space-times, particularly to linear plane gravitational waves in the cross mode, and we do not have any counterexample, it has not been shown that it is always possible to carry this step out. That is why we present an implementation of the principle of equivalence, which we call the principle of rigid equivalence, in the form of a conjecture. 

We have succeeded in finding Fermi rigid coordinates for a type of linear gravitational wave, and this could be an interesting and useful alternative when trying to invest Fermi normal coordinates with physical meaning. It seems that, using Fermi rigid coordinates, we are free to interpret space rigid coordinates as those that label, at least approximately, points on a body made of rigid material, in the sense that the cohesive forces of this material are much more intense than the tensions caused by the passage of the gravitational wave. This feature may be interesting for those experimenters who need good correspondence between the mathematical symbols used in the theory and the laboratory tools that they use to design experimental devices. Whatever the case, we believe that the rigid formulation proposed herein can assist in clarifying some open questions related to the correspondence between the mathematical formulation of general relativity and how to perform laboratory experiments.

The Fermi coordinate condition does not determine which ones are the ``good coordinates''. It would be interesting to study the relationship between the different proposals that incorporate the rigidity in relativity \cite{Bona83, Bel90, Llosa04, Coll07} with that presented here and with the Fermi condition. This study is beyond the scope of this work and we hope to deal with it in detail in the future.

\subsection*{Acknowledgments}
I want to thank Alfred Molina for carefully reading a previous draft of the paper and providing useful criticism that led to improvements; and Llu\'{i}s Bel, without whose inspiration and encouragement, hardly any of this series of papers would have occurred to me.

\end{document}